\documentclass[11pt,a4paper]{article}

\usepackage[utf8]{inputenc}
\usepackage[T1]{fontenc}
\usepackage{amsmath,amssymb}
\usepackage{graphicx}
\usepackage{xcolor}
\usepackage{colortbl}
\usepackage{steinmetz}
\usepackage{authblk}
\usepackage{geometry}
\usepackage{hyperref}
\usepackage{lineno}

\graphicspath{{./images/}}

\hypersetup{
    colorlinks=true,
    linkcolor=blue,
    citecolor=blue,
    urlcolor=blue
}

\begin{document}

\title{
Dispersive-phonon-driven room-temperature Ni$^{1+}$--Ni$^{2+}$
polaron hopping in spin--charge coupled rutile niobate
}

\author[1]{Gourab Roy}
\author[1]{Mohit Kumar}
\author[1]{Sayan Ghosh}
\author[1]{Ekta Kushwaha}
\author[2]{Manh Duc Le}
\author[2,3]{Devashibhai T. Adroja}
\author[1]{Tathamay Basu}

\affil[1]{
Rajiv Gandhi Institute of Petroleum Technology,
Jais, Amethi 229304, India
}

\affil[2]{
ISIS Neutron and Muon Source, STFC,
Rutherford Appleton Laboratory,
Harwell Campus, Didcot, Oxfordshire OX11 0QX, United Kingdom
}

\affil[3]{
Physics Department, Highly Correlated Matter Research Group,
University of Johannesburg,
Auckland Park 2006, South Africa
}

\date{}

\maketitle

\abstract{Understanding how lattice dynamics mediate polaron hopping is essential for designing multifunctional correlated oxides. Here, we demonstrate room-temperature dispersive phonon excitations and elucidate the $\mathrm{Ni}^{1+}$--$\mathrm{Ni}^{2+}$ polaron-hopping mechanism and the presence of rare spin-charge-phonon coupling even in a magnetically short-range-ordered state in rutile niobate, a rare room-temperature magnetodielectric system. We reveal room-temperature dispersive phonon excitations using inelastic neutron scattering (INS), complemented by machine-learning-based phonon calculations, to establish the microscopic origin of the polaron-hopping mechanism. Experimental evidence of dispersive phonon-driven polaron hopping is scarce. INS measurements show significant dispersive phonon excitations at 21, 33, and 47 meV, implying collective lattice dynamics that enable delocalized polaron propagation via coupled charge-spin-phonon interactions. Dispersive phonons couple to charge carriers and promote correlated $\mathrm{NiO_6}$ lattice distortions, facilitating delocalized polaron hopping. Low-energy magnetic excitations at 4 and 8 meV indicate the presence of local short-range magnetic correlations or spin-orbit-coupling-induced anisotropy in deformed $\mathrm{NiO_6}$ octahedra, which are thoroughly discussed. Machine-learning phonon calculations replicate the experimentally observed phonon excitations and demonstrate lattice instability, which is compatible with dynamic local distortions caused by polaron production. These findings provide microscopic evidence for a coupled charge-spin-phonon mechanism that mediates polaron hopping in rutile oxide systems.  }





\maketitle


\section{Introduction}

Polarons arising from strong charge--lattice coupling play a central role in determining the electronic and magnetic properties of correlated oxides ~\cite{Shneyder2020, Mirjolet2021, Franchini2021, Choi2023}. In mixed-valence transition-metal systems, thermally activated polaron hopping is strongly influenced by the coupling among charge, spin, and lattice degrees of freedom ~\cite{Verdi2017, Hasan2026}. The nature of the phonon modes is essential to polaron dynamics; dispersive (Debye-like) phonons promote extended lattice distortions and increased polaron mobility, whereas non-dispersive (Einstein-like) phonons localise the distortion around individual sites, favouring small-polaron formation and thermally activated hopping ~\cite{ Franchini2021, Zhang2023}. Such polaronic mechanisms have been explored in oxide materials, including manganites ~\cite{Raiser2017, Weber2021}, titanates ~\cite{Choi2015, Boyd2025}, cuprates ~\cite{ Velasco2022}, nickelates ~\cite{ Shamblin2018}, and mixed-valence transition-metal compounds ~\cite{Verdi2017, Hasan2026}. Polaron hopping has been extensively studied using Holstein and Frohlich models ~\cite{Holstein1959, Frohlich1954}, density functional theory (DFT) ~\cite{Sio2019, Meggiolaro2020}, dynamical mean-field theory (DMFT) ~\cite{Fratini2003}, and electron-phonon coupling frameworks to better understand the relationship between charge localisation ~\cite{Sio2019}, lattice distortion, spin interactions, and carrier mobility in correlated oxides. However, experimentally demonstrating the microscopic polaron-hopping mechanism remains challenging. Polaron hopping, arising from strong electron--phonon coupling, plays a pivotal role in governing charge transport and the coupled magnetic, dielectric, optical, and thermoelectric properties of a wide range of correlated oxides and functional materials. Understanding its microscopic mechanism is therefore essential for the rational design of next-generation magnetic, semiconductor, energy-storage, photocatalytic, and multifunctional electronic materials~\cite{Buizza2021, Ren2024, Guo2026, Gomes2026}.

Notably, the Niobate family shows versatile potential applications across various aspects, such as battery materials for energy storage, photoluminescence, photocatalysis, magnetism, and magnetodielectric coupling, depending on its crystal structure, which varies with sample preparation techniques ~\cite{, Kampfe2016, Kong2020, Wang2020, Waqar2022}. Among such materials, NiNb$_2$O$_6$ has attracted considerable interest owing to its wide structural variety with corresponding a unique magnetic behaviour. Previous neutron diffraction and $\mu$SR studies demonstrated that the $\beta$ polymorph of NiNb$_2$O$_6$ ($P4_2/n$) \cite{Munsie2017}, in comparison with the columbite phase (Pbcn) \cite{Maruthi2023}, exhibits a distinct low-dimensional $S=1$ antiferromagnetic spin-chain ground state with stronger interchain exchange interactions and a higher magnetic ordering temperature, illustrating the diverse magnetic phenomena arising from structural polymorphism in the NiNb$_2$O$_6$ family.

Very recently, the rutile nickel niobate compound \(\mathrm{NiNb_2O_6}\) was reported to exhibit a room-temperature magnetodielectric effect, in which polaron hopping between mixed-valence \(\mathrm{Ni}^{1+}\) and \(\mathrm{Ni}^{2+}\) ions is proposed to play a vital role ~\cite{Wang2020, Wang2021}. However, despite previous results indicating polaron-mediated transport, the microscopic origin of polaron formation and hopping in the context of charge transfer, lattice dynamics, and magnetic excitations remains only partially understood experimentally, leaving a gap for investigation. The correlated interactions in rutile \(\mathrm{NiNb_2O_6}\) are an ideal platform to investigate the \(\mathrm{Ni}^{1+}\)--\(\mathrm{Ni}^{2+}\) polaron-hopping mechanism through exploring the coupled evolution of charge transfer, lattice dynamics, local spin fluctuations, and dispersive phonon excitations. However, though not been fully explored yet. Here, we demonstrate a detailed microscopic picture of polaron hopping using inelastic neutron scattering measurements, accompanied by theoretical calculations. Neutron scattering techniques are unique and high-level tools to explore the ground state of materials and thereby fundamental mechanisms. Further, these results are supported by machine-learning-based phonon calculations, which have emerged as an important modern approach for investigating lattice dynamics and correlated excitations in materials science.

\section{Experimental Details}

Polycrystalline rutile NiNb$_2$O$_6$ (50 wt.\% NiNb$_2$O$_6$ + 50 wt.\% TiO$_2$) was synthesised by the conventional solid-state reaction method ~\cite{Wang2020, Wang2021}. Stoichiometric amounts of NiO, Nb$_2$O$_5$, and TiO$_2$ powders were thoroughly mixed and ground using an agate mortar and pestle. Then the mixture was formed into a pellet by applying pressure. The pellet was heated at $1500\,^{\circ}$C for 10~h, followed by quenching in air to stabilise the rutile phase. High-resolution synchrotron powder X-ray diffraction (SXRD) data were collected at room temperature at the XRD beamline of the RRCAT, Indore, using an incident wavelength of $\lambda = 0.82298$~\AA. Rietveld refinement of the SXRD data confirmed the formation of the desired single-phase rutile NiNb$_2$O$_6$ crystallising in the tetragonal space group $P4_2/mnm$ (see Supplemental Material Fig.~S1). Inelastic neutron scattering (INS) measurements were performed on the MARI time-of-flight spectrometer at the ISIS Neutron and Muon Source, UK (Experiment RB2590216) ~\cite{Basu2025}, using approximately 3~g of powder sealed in an aluminium can. The measurements were carried out in repetition rate multiplication (RRM) mode with a Gd-Fermi chopper operating at 400~Hz. The primary incident neutron energy was $E_i = 180$~meV, with additional incident energies of 100, 29.7, 22.9, and 11.7~meV. INS spectra were collected at temperatures of 6, 100, and 300~K. The neutron scattering data were reduced and analysed using the Mantid and DAVE software packages~\cite{Arnold2014, Azuah2009}. Phonon calculations were performed using the INSPIRED software package~\cite{Yang2024}, which employs pre-trained machine-learning force fields (MLFFs) developed within the MatterSim framework to simulate the lattice dynamics and neutron-weighted phonon spectra.

\vspace{0.5em}

\begin{figure*}[t]
    \centering
    \includegraphics[width=0.8\textwidth]{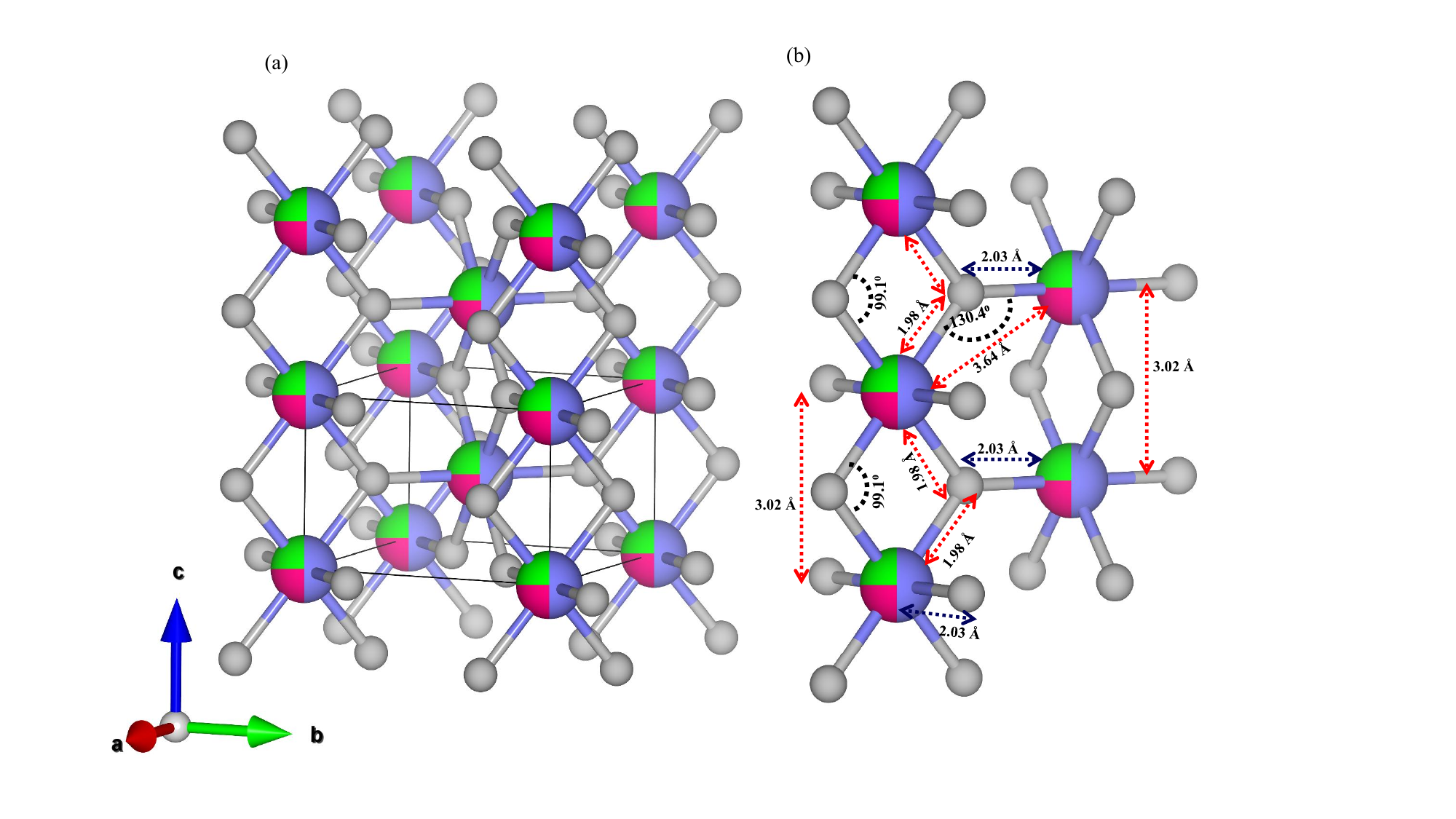}
    \caption{(a) Crystal structure of $\mathrm{NiNb_2O_6}$ crystallizing in the tetragonal $P4_2/mnm$ space group. Violet and green spheres represent Nb and Ni atoms, respectively, while grey spheres denote oxygen atoms. (b) Enlarged view of the local octahedral environment showing the $M$--O bond lengths and $M$--O--$M$ bond angles.}
    \label{fig:crystal_structure}
\end{figure*}

\section{Results and Discussions}

\subsection{Crystal Structure Analysis}

Rutile \(\mathrm{NiNb_2O_6}\) crystallises in a single-phase tetragonal structure with space group \(P4_2/mnm\) ~\cite{Wang2020}, as confirmed by Rietveld refinement of the XRD data (see Supplemental Material Fig.~S1  ~\cite{SupplementalMaterial}). The corresponding Wyckoff positions and occupancies are listed in Table~S1 of the Supplemental Material ~\cite{SupplementalMaterial}. The crystal structure is depicted in Fig. 1(a). Ni, Nb, and Ti ions share the same Wyckoff position \((0,0,0)\). Fig. $1$(b) shows detailed bond lengths and angles to explain the crystallographic genesis of the polaron hopping mechanism. For \(\mathrm{NiO_6}\) octahedra, the minimum Ni--O bond length is around \(1.87~\text{\AA}\), the maximum is \(2.46~\text{\AA}\), and the mean bond length is around \(2.07~\text{\AA}\) ~\cite{Gagne2020}. In this system, all octahedra (containing Ni, Nb, Ti) are crystallographically equivalent, consisting of two longer bonds (\(\sim 2.03~\text{\AA}\) lying in the \(xy\)-plane and four comparatively shorter bonds (\(\sim 1.98~\text{\AA}\)). Short bonding, such as \(1.87~\text{\AA}\), can enhance the overlap between Ni \(3d\) orbitals and O \(2p\) orbitals within the \(xy\)-plane, thereby increasing Coulomb interactions and favoring Jahn--Teller (JT) distortion for \(\mathrm{Ni^{1+}}\) (\(3d^9\)), analogous to the well-known case of \(\mathrm{Cu^{2+}}\) ~\cite{ Halcrow2013}. The suggested JT distortion can lift the degeneracy of the \(d_{z^2}\) and \(d_{x^2-y^2}\) orbitals (as discussed in further details in the Conclusion and the orbital splitting is shown in Fig. 5(b)), comparable to the z-in distortion. The shorter apical bond lengths (\(\sim 1.98~\text{\AA}\)) may cause the \(d_{z^2}\) orbital energy to exceed the \(d_{x^2-y^2}\) level.  Here the observed Ni-O-Ni bond angles of \(99.11^\circ\) and \(130.44^\circ\) deviate significantly from \(180^\circ\). According to the Goodenough--Kanamori--Anderson (GKA) principles ~\cite{Goodenough1955}, such notably bent superexchange paths result in weak antiferromagnetic interactions and thus weak magnetic exchange coupling.

\subsection{Inelastic Neutron Scattering}  

\begin{figure*}[t]
    \centering
    \includegraphics[width=.8\textwidth]{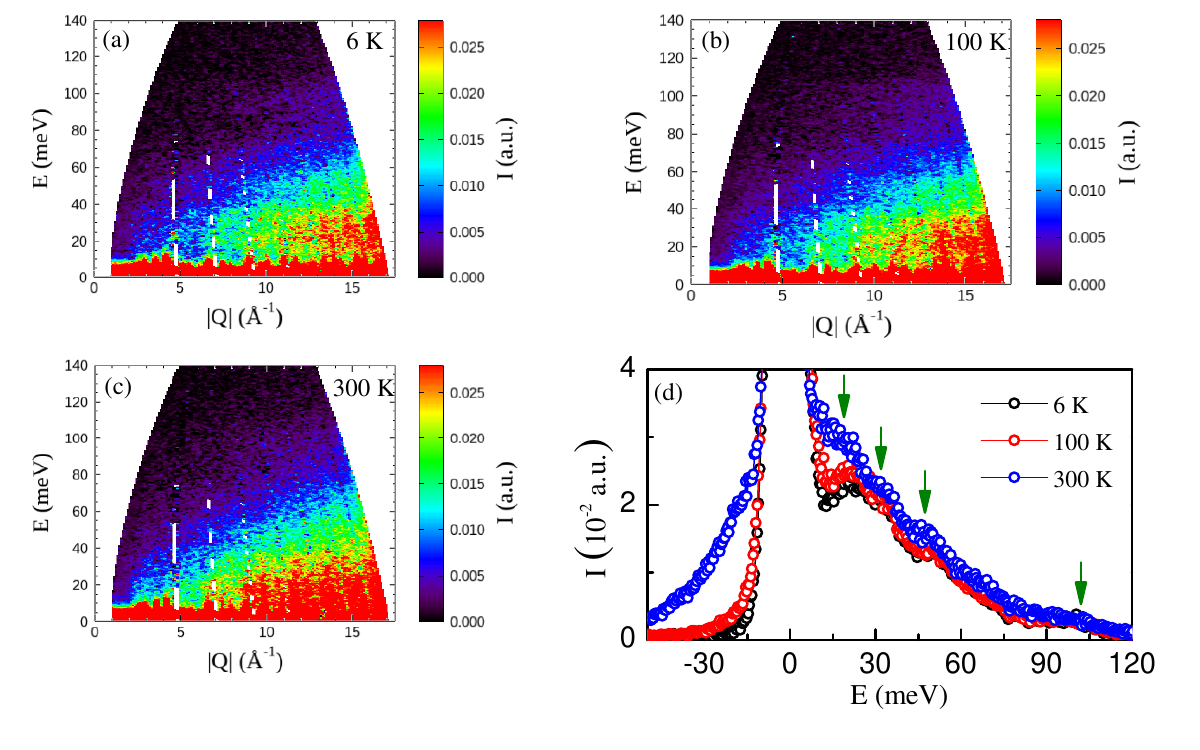}
    \caption{(a-c) Color contour plots for $E_i = 180$ meV, revealing energy excitations at $6$, $100$, and $300$ K, respectively. (d) Comparison of phonon excitations (Intensity(I) vs. Energy Transfer) at $6$, $100$ and $300$ K in the higher $Q$ region $8$ to $16~\text{\AA}^{-1}$.}
    \label{fig1}
\end{figure*}

\begin{figure*}[t]
    \centering
    \includegraphics[width=1\textwidth]{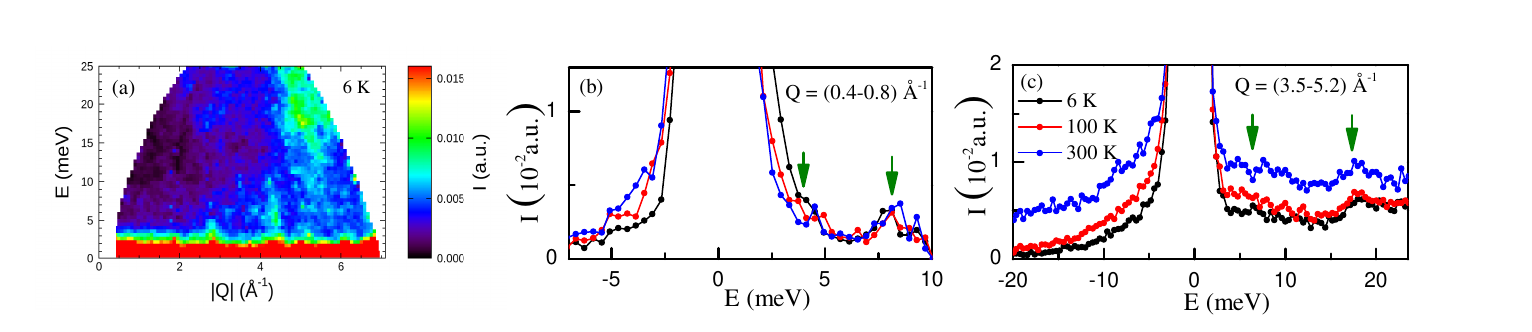}
    \caption{
(a) Color contour plots for $E_i = 29.7\,\mathrm{meV}$, showing energy excitations at $6\,\mathrm{K}$. (b) Comparison of magnetic excitations at $6\,\mathrm{K}$, $100\,\mathrm{K}$, and $300\,\mathrm{K}$ in the low-$Q$ region ($0.4$--$0.8\,\text{\AA}^{-1}$). (c) Comparison of phonon excitations at $6\,\mathrm{K}$, $100\,\mathrm{K}$, and $300\,\mathrm{K}$ in the high-$Q$ region ($3.5$--$5.2\,\text{\AA}^{-1}$).
}
    \label{fig2}
\end{figure*}

Understanding the nature of magnetic and phonon excitation is essential for investigating the origin of polaron hopping. Therefore, we performed inelastic neutron scattering (INS) measurements with $E_i = 180\,\mathrm{meV}$ over a range of temperatures ($6$, $100$, and $300\,\mathrm{K}$) ~\cite{Basu2025}, as illustrated in the colour contour plots in Fig. 2(a)--(c). In the high-$Q$ region, the intensity ($I$) increases with momentum transfer ($Q$), particularly in the range of $8$--$16\,\text{\AA}^{-1}$ (see Fig. S2 in Supplemental Material ~\cite{SupplementalMaterial}). From the I vs Q curve the dispersive nature of the phonon is evident too. No significant excitation is observed in the low-$Q$ region. Instead, strong excitations appear in the high-$Q$ region, which become more pronounced with increasing temperature. These behaviours indicate that these excitations originate from phonon contributions. Fig. ~2(d) presents the corresponding one-dimensional ($1$D) plot, showing phonon excitations over a wide energy range, with prominent features around $21$, $33$, $47$, and $100\,\mathrm{meV}$. The $100$ meV phonon modes are found to be very weak compared to other phonon excitations. Interestingly, it is observed that the phonon feature is in a dispersive wave-like nature instead of a flat line excitation (see Fig.$2$(a)-(c)). This particular phonon feature shows a significant energy dependence on Q, and hence arises from the dispersive nature of phonon modes. With increasing temperature, particularly at 300~K (see Fig. $2$(d)), the phonon response exhibits significant broadening, redistribution of spectral weight, and enhanced diffuse intensity over a wide energy range. Such anomalous temperature-dependent phonon modification indicates effective electron-phonon coupling, consistent with behaviour observed in correlated electron systems ~\cite{Kurzhals2022}.

\vspace{0.5em}

In a backdrop of dispersive (Debye-like) phonons, the lattice behaves as a coupled system, with energy based on relative displacements between nearby atoms rather than individual site displacement ~\cite{Blackham2025}. As a result, here phonon dispersion has a significant impact on polaron properties by spreading the lattice deformation, in contrast to a localised polaronic state caused by nondispersive phonons (Einstein-like). In a polaronic system, phonons influence how the lattice responds when an electron hops across sites, such as the $\mathrm{Ni}^{1+}$ to $\mathrm{Ni}^{2+}$ transfer in tightly coupled $\mathrm{NiO}_6$ octahedra. When an extra electron occupies a site (as in $\mathrm{Ni}^{1+}$), it causes a local distortion that tightly binds it, resulting in a polaron. Because of this, a sharp, localised distortion at one $\mathrm{NiO}_6$ octahedron would create large differences with its neighbours and cost a lot of energy. To minimise this energy, the system naturally prefers a smooth distortion pattern, where nearby sites are displaced similarly. Therefore, when an electron localises near a $\mathrm{Ni}^{1+}$ site, the distortion spreads over several $\mathrm{NiO}_6$ units, resulting in a delocalized polaron due to the presence of dispersive phonon modes as further demonstrated in the Conclusion in Fig.~5(a). This extended deformation enables continuous charge transfer between $\mathrm{Ni}^{1+}$ and $\mathrm{Ni}^{2+}$ sites, exhibiting polaron hopping with lower energy cost. The lattice distortion evolves collectively and spreads from one site to another site by charge transfer and collective interaction of dispersive phonon modes. As the temperature increases, the phonon modes increase, facilitating pronounced polaron hopping at higher temperatures as the corresponding electron-phonon interaction is increased.

\vspace{0.5em}

\begin{figure*}[t]
    \centering
    \includegraphics[width=.8\textwidth]{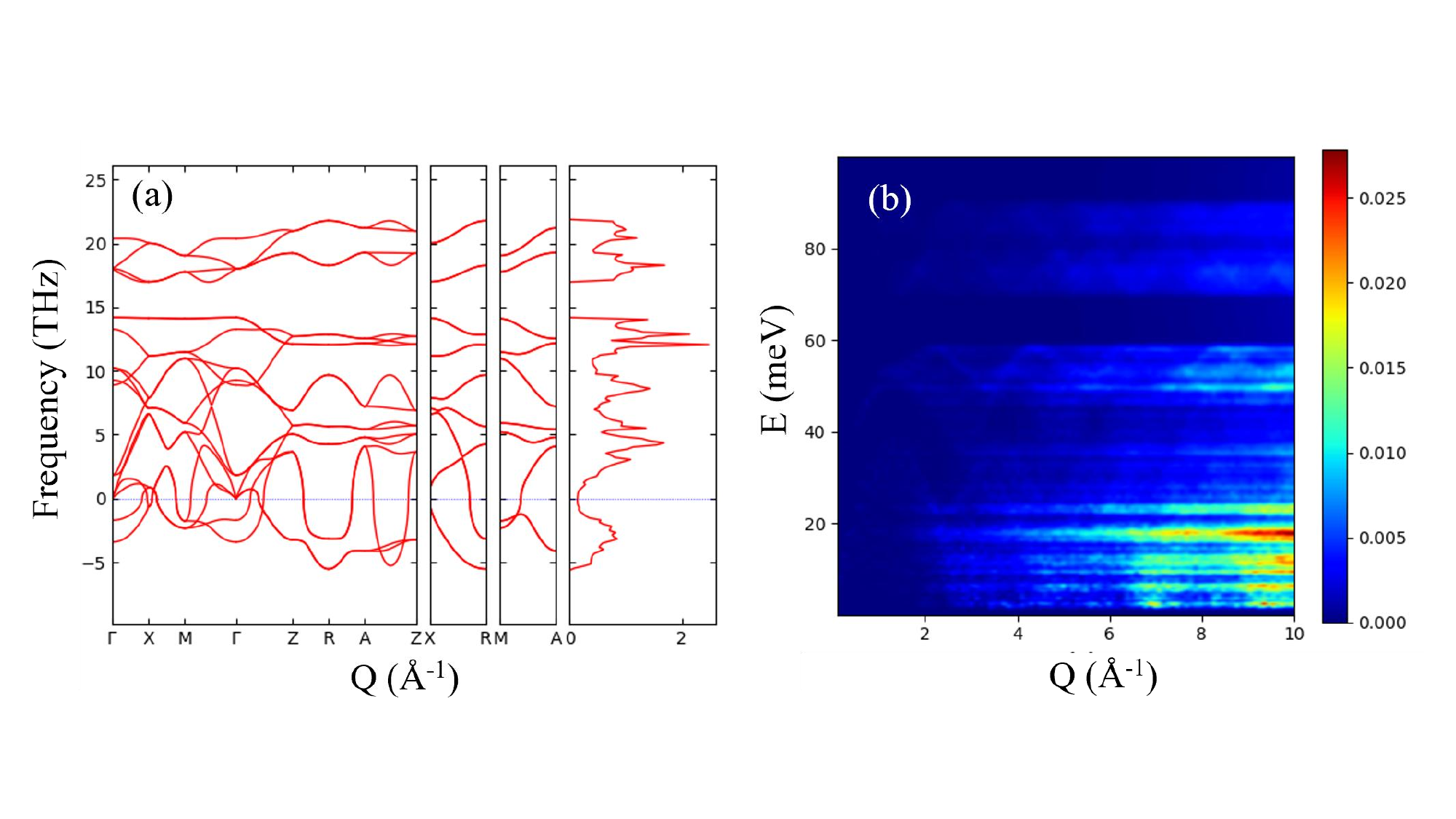}
    \caption{
    (a) Calculated phonon modes based upon pre-trained machine learning force fields. The vertical axis represents the phonon energy in units of THz. Panels indicate wave-vector dependence in reciprocal space using standard reciprocal-space notation. (b) Calculated phonon scattering intensity at 300 K.
    }
    \label{fig1}
\end{figure*}

\vspace{0.5em}

To study low energy excitation, we performed with $E_i = 29.7\,\mathrm{meV}$ at temperatures $6$, $100$, and $300\,\mathrm{K}$ ~\cite{Basu2025}. The corresponding colour contour plot, as shown in Fig.~3(a)--(c), shows excitation at the high $Q$ region $3.5$--$5.2\,\text{\AA}^{-1}$, where the intensity of excitation increases with $Q$ (see Fig. S3 in Supplemental Material), which depicts phonon excitations. Although these phonon features are not as strong as reported in Fig.~2, the 1D plot in Fig.~3(c), showing intensity ($I$) vs energy transfer ($E$), reveals phonon excitations at about $7$ and $17\,\mathrm{meV}$. The colour contour map cannot definitively determine whether these modes are dispersive or non-dispersive. As a result, non-dispersive phonon modes may exist along with dispersive modes, allowing for localised polaron production, with hopping becoming more active at higher temperatures as thermal energy exceeds the activation barrier. Although the latter section in the machine learning simulation, we have shown that almost all the phonon modes are of a dispersive nature. Overall, in this rutile $\mathrm{NiNb}_2\mathrm{O}_6$, experimentally it is observed that the major phonon contributions at $21$, $33$, $47\,\mathrm{meV}$ are dispersive in nature, which could mostly assist in polaron formation and hopping mechanisms.

\vspace{0.5em}

In the low Q region $0.4$--$0.8\,\text{\AA}^{-1}$ of Fig. $3$(a), it is observed that I decay with Q, (see Fig. S4 in Supplemental Material ~\cite{SupplementalMaterial}), where two small excitations are observed around $4$ and $8$ meV, as shown in $1$D plot Fig. 3(b). From this plot, although the room temperature magnetic excitation is present, the temperature dependence with $6$-$300$ K is not clearly understood, as the intensity of excitation is low. So this excitation must come from a magnetic origin. Although the system is disordered and lacks long-range order (LRO), these excitations may originate from short-range magnetic correlations or spin–orbit coupling (SOC) effects, which are discussed in detail.

Despite the absence of long-range order (LRO), these magnetic patterns may be the result of short-range magnetic correlations that persist from $6\,\mathrm{K}$ to $300\,\mathrm{K}$. Polaron hopping occurs at higher temperatures via the $\mathrm{Ni}^{1+}$--O--$\mathrm{Ni}^{2+}$ pathway, leading in a spin state change from $S=\frac{1}{2}$ to $S=1$. Magnetic correlations could be maintained at room temperature by activating the $\mathrm{Ni}^{1+}$--O--$\mathrm{Ni}^{2+}$ channel, which is driven by charge transfer and spin-state changes during the polaron hopping process. Since the hopping occurs between $\mathrm{Ni}^{1+}$ and $\mathrm{Ni}^{2+}$ sites, the spin correlations are expected to be short-ranged and dynamic, moving locally through the system rather than generating static long-range magnetic order.

Another strong possibility is that the local origin SOC of ZFS can break the $\mathrm{Ni}^{2+}$ $S=1$ state, as SOC cannot divide a pure $\mathrm{Ni}^{1+}$ $S=1/2$ doublet in zero field ~\cite{Boca2004, Sharma1966}. The zero-field splitting Hamiltonian for $\mathrm{Ni}^{2+}$ with spin $S=1$ is as follows:

\[H = D S_z^2 + E(S_x^2-S_y^2) \].

Using the basis states \[|1\rangle,\quad |0\rangle,\quad |-1\rangle\] the spin operator $S_z$ can be written as

\[
DS_z^2=\begin{pmatrix}
D&0&0\\
0&0&0\\
0&0&D
\end{pmatrix}
\]

Using the ladder operators,
\[
S_x^2-S_y^2=\frac12(S_+^2+S_-^2)
\]

For \(S=1\),
\[
S_+^2|-1\rangle=2|1\rangle,\quad
S_-^2|1\rangle=2|-1\rangle
\]

which gives
\[
E(S_x^2-S_y^2)=\begin{pmatrix}
0&0&E\\
0&0&0\\
E&0&0
\end{pmatrix}
\]

The total Hamiltonian matrix becomes
\[
H=\begin{pmatrix}
D&0&E\\
0&0&0\\
E&0&D
\end{pmatrix}
\]

The eigenvalues are obtained from
\[
\det(H-\lambda I)=\begin{vmatrix}
D-\lambda&0&E\\
0&-\lambda&0\\
E&0&D-\lambda
\end{vmatrix}=0
\]

which gives
\[
-\lambda\left[(D-\lambda)^2-E^2\right]=0
\]

Hence, the \(S=1\) triplet splits into three levels:
\[
E_1=0,\quad E_2=D+E,\quad E_3=D-E
\]

\begin{figure*}[t]
    \centering
    \includegraphics[width=.8\textwidth]{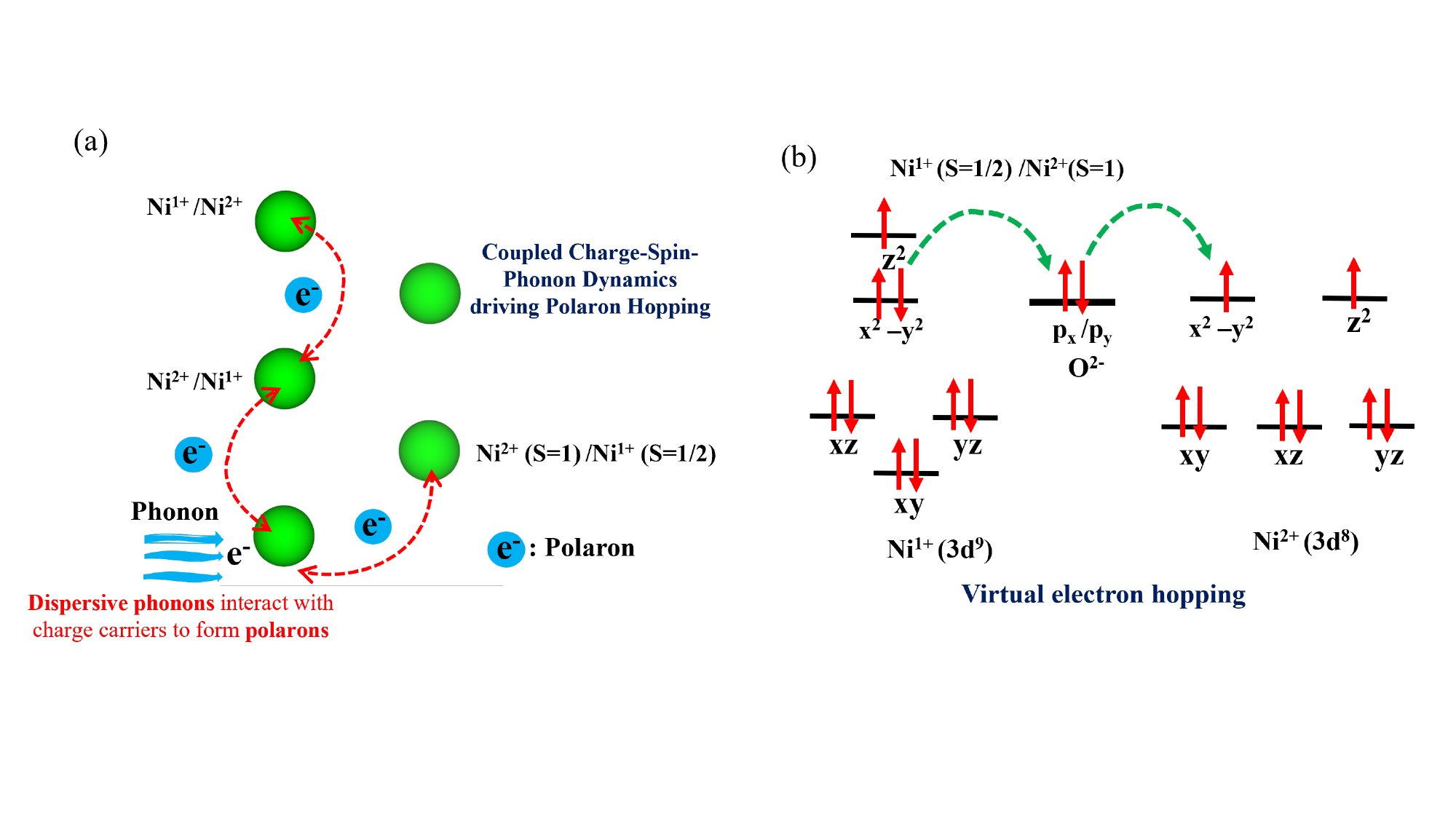}
    \caption
{ 
(a) Schematic illustration of polaron hopping dynamics between $\mathrm{Ni}^{1+}$ and $\mathrm{Ni}^{2+}$ sites mediated by coupled charge--spin--phonon interactions. The hopping electron locally distorts neighbouring $\mathrm{NiO_6}$ octahedra, facilitating distortion propagation during charge transport. 
(b) Illustration of virtual electron hopping between $\mathrm{Ni}^{1+}$ and $\mathrm{Ni}^{2+}$ sites, which assists the polaron hopping mechanism through spin-dependent electronic exchange pathways.}

    \label{fig1}
\end{figure*}

Thus, the transitions from the ground state $|0\rangle$ to the mixed excited states 
$\frac{1}{\sqrt2}(|+1\rangle-|-1\rangle)$ and $\frac{1}{\sqrt2}(|+1\rangle+|-1\rangle)$, correspond to the observed magnetic excitations at around $4$ and $8\,\mathrm{meV}$, respectively. $DS_z^2$ represents axial anisotropy, which divides the $S=1$ triplet into $|0\rangle$ and $|\pm1\rangle$ levels, while $E(S_x^2-S_y^2)$ represents transverse anisotropy, which mixes the $|+1\rangle$ and $|-1\rangle$ states due to local lattice distortion. These terms work together to produce low-energy magnetic excitations due to CEF and SOC effects. The observed excitations at $4$ and $8\,\mathrm{meV}$ indicate substantial axial anisotropy with finite transverse distortion coming from CEF and SOC effects in distorted $\mathrm{NiO}_6$ octahedra. The estimated anisotropy parameters are $D=6\,\mathrm{meV}$ and $E=2\,\mathrm{meV}$. Therefore, if these excitations originate from single-ion anisotropy of $\mathrm{Ni}^{2+}$ ions, the relatively large anisotropy ratio $(E/D \approx 0.33)$ indicates significant transverse (rhombic) distortion of the local $\mathrm{NiO_6}$ environment, which may influence the local lattice dynamics and indirectly assist the formation and hopping of polarons. The appearance of a sharp EPR signal ($g \approx 2.076$) and a broad anisotropic band ($g \approx 2.816$) suggests the presence of single-ion anisotropy caused by crystal-field effects and spin-orbit coupling (SOC) ~\cite{Wang2020}. This anisotropy causes zero-field splitting that matches the low-energy magnetic excitations ($\sim 4$ and $\sim 8\,\mathrm{meV}$) observed in INS. These magnetic excitations could be caused by short-range magnetic correlations and SOC effects, or by a combination of single-ion anisotropy of $\mathrm{Ni}^{2+}$, short-range magnetic correlations, and polaron-induced local distortions, resulting in coupled spin-orbital-lattice excitations.

\subsection{Machine Learning Models for Phonon Excitation}

We calculated the phonons across the whole $|$Q$|$-range to distinguish them from the magnetic signals observed in the INS data in a machine learning framework, which employs pre-trained machine-learning force fields (MLFFs) derived from the MatterSim modelling package ~\cite{Cheng2023, Roy2026}. We used the 300 K crystal structure and optimised it. Fig. $4$(a) illustrates the computed phonon spectrum, which covers up to 25 THz. This calculation also took into account numerous phonon processes and the temperature-dependent thermal population of phonons. The estimated phonon dispersion exhibits negative frequencies (or imaginary phonon modes) as shown in Fig. $4$(a), indicating dynamic disorder in the simulated crystal structure of $\mathrm{NiNb_2O_6}$. Similar imaginary phonon modes have also been reported in earlier studies to describe dynamic structural instability associated with lattice distortions ~\cite{Manti2023, PhysRevB.106.195407}. In rutile $\mathrm{NiNb_2O_6}$, such instability is related to local structural distortions caused by $\mathrm{NiO_6}$ octahedra, possible Jahn--Teller-like events, or dynamic displacement of oxygen molecules, which has already been discussed. These distortions disrupt crystal symmetry while stabilising the structure via local phonon softening. Thus, the presence of negative phonon density of states (DOS) or imaginary frequencies implies that the undistorted high-symmetry structure is dynamically unstable and drifts toward a lower-energy distorted ground state. This observation indicates the potential of substantial spin--phonon and electron--phonon coupling associated with polaron hopping in rutile $\mathrm{NiNb_2O_6}$. Local lattice distortions accompany the $\mathrm{Ni}^{1+}/\mathrm{Ni}^{2+}$ charge fluctuations.

The simulated intensity (Fig.$4(b)$) shows that phonon scattering dominates the $|$Q$|$ region $5$ to $10~\text{\AA}^{-1}$, consistent with the measured phonon modes. The low-energy excitations that appear broad and unresolved in the experimental INS data below $40 meV$ (see Fig. 2(d)) are due to both the closely spaced modes and the instrumental energy resolution in this energy-transfer regime. But here in simulated phonon spectra, each mode can be visualised properly.  This calculation shows that the excitations observed in the INS data at 21, 33, and 47~meV exhibit a dispersive nature, while the remaining excitations at 7, 17, and 100~meV appear with weak intensity but also look like a dispersive wave-like nature. All of these excitations that were observed in INS are present, as shown in the colour plot of Fig.~4(b). These calculated energies match very well with the excitations observed in the experimental spectra (see Fig. $2$), confirming both the accuracy of the phonon calculation and its consistency with the experimental INS results. Such a clear agreement between the INS experimental phonon spectra and the theoretical MLFF simulated phonon spectra is also observed in another transition metal-based oxide system. ML-based force fields provide an efficient route to dense sampling of the phonon spectrum and enable direct comparison with INS intensity over broad Q–E ranges, without implying that such analyses would be unattainable using conventional approaches.

\section{Conclusion}

In this study, we demonstrate that polaron-hopping dynamics between $\mathrm{Ni}^{1+}$ and $\mathrm{Ni}^{2+}$ sites are mediated by coupled charge--spin--phonon interactions as shown in Fig. $5$(a). Polaron hopping originates through charge transfer between neighbouring Ni ions via virtual electron hopping mediated by oxygen, as seen in Fig. $5$(b). During the hopping process, the valence states of $\mathrm{Ni^{1+}}$ and $\mathrm{Ni^{2+}}$ interchange dynamically, followed by local lattice distortions that spread along the charge transfer process. The local lattice distortion traps an extra electron around the $\mathrm{Ni^{1+}}$ site, resulting in a polaronic state due to strong electron-phonon interactions as demonstrated in Fig. $5$(a). Inelastic neutron scattering (INS) studies show the presence of electron-phonon interaction and the dispersive phonon modes that are responsible to the hopping mechanism. The machine learning framework addresses dynamic distortions. The connected lattice oscillations assist the partial delocalisation and transmission of the polaronic state throughout the lattice, indicating an intrinsic charge-spin-phonon coupled mechanism that governs the polaron dynamics in rutile $\mathrm{NiNb_2O_6}$.

\section{Acknowledgements}

TB greatly acknowledges the Science and Engineering Research Board (SERB) (Project No.: SRG/2022/000044). T.B.\ thanks the Science and Technology Facilities Council (STFC), UK, for providing inelastic neutron scattering beam time on the Mari (proposal no.\ RB2590216) at the ISIS Neutron and Muon Facility. TB thanks the Oak Ridge National Laboratory (ORNL) for providing access to the INSPIRED software facility for machine-learning-based phonon calculations. T.B. thanks Dr. Archna Sagdeo from RRCAT for his assistance with the synchrotron powder X-ray diffraction experiments. DTA would like to thank the EPSRC UK for the funding (Grant No. EP/W00562X/1). GR acknowledge the Raman-Charpak Fellowship by CEFIPRA.

\section{Conflicts of Interest}

The authors declare no conflicts of interest.



\bibliographystyle{unsrt}
\bibliography{References}

@article{Shneyder2020,
  author = {Shneyder, E. I. and Nikolaev, S. V. and Zotova, M. V. and Kaldin, R. A. and Ovchinnikov, S. G.},
  title = {Polaron transformations in the realistic model of the strongly correlated electron system},
  journal = {Phys. Rev. B},
  volume = {101},
  pages = {235114},
  year = {2020},
  doi = {10.1103/PhysRevB.101.235114}
}

@article{Franchini2021,
  author = {Franchini, C. and Reticcioli, M. and Setvin, M. and Diebold, U.},
  title = {Polarons in materials},
  journal = {Nat. Rev. Mater.},
  volume = {6},
  pages = {560--586},
  year = {2021},
  doi = {10.1038/s41578-021-00278-8}
}

@article{Choi2023,
  author  = {Choi, I. H. and Jeong, S. G. and Min, T. and Lee, J. and Choi, W. S. and Lee, J. S.},
  title   = {Giant Enhancement of Electron--Phonon Coupling in Dimensionality-Controlled SrRuO$_3$ Heterostructures},
  journal = {Advanced Science},
  volume  = {10},
  number  = {16},
  pages   = {2300012},
  year    = {2023},
  doi     = {10.1002/advs.202300012}
}

@article{Mirjolet2021,
  author  = {Mirjolet, M. and Rivadulla, F. and Marsik, P. and Borisov, V. and Valent{\'i}, R. and Fontcuberta, J.},
  title   = {Electron--phonon coupling and electron--phonon scattering in SrVO$_3$},
  journal = {Advanced Science},
  volume  = {8},
  number  = {15},
  pages   = {2004207},
  year    = {2021},
  doi     = {10.1002/advs.202004207}
}

@article{Hasan2026,
  author  = {Hasan, Zubia and Pan, Grace A. and LaBollita, Harrison and
             Kaczmarek, Austin and Sung, Suk Hyun and Sharma, Shekhar and
             Balakrishnan, Purnima P. and Mercer, Edward and Bhartiya, Vivek and
             N'Diaye, Alpha T. and Salman, Zaher and Prokscha, Thomas and
             Suter, Andreas and Grutter, Alexander J. and
             Garcia-Fernandez, Mirian and Zhou, Ke-Jin and
             Pelliciari, Jonathan and Bisogni, Valentina and
             El Baggari, Ismail and Schlom, Darrell G. and
             Barone, Matthew R. and Brooks, Charles M. and
             Nowack, Katja C. and Botana, Antia S. and Faeth, Brendan D. and
             de la Torre, Alberto and Mundy, Julia A.},
  title   = {Unconventional polaronic ground state in superconducting LiTi$_2$O$_4$},
  journal = {Nature Communications},
  volume  = {17},
  number  = {1},
  pages   = {1303},
  year    = {2026},
  doi     = {10.1038/s41467-025-68068-7}
}

@article{Verdi2017,
  author = {Verdi, C. and Caruso, F. and Giustino, F.},
  title = {Origin of the crossover from polarons to Fermi liquids in transition metal oxides},
  journal = {Nat. Commun.},
  volume = {8},
  pages = {15769},
  year = {2017},
  doi = {10.1038/ncomms15769}
}

@article{Zhang2023,
  author = {Zhang, C.},
  title = {Effect of dispersive optical phonons on the properties of the bond Su-Schrieffer-Heeger polaron},
  journal = {Phys. Rev. B},
  volume = {108},
  pages = {075156},
  year = {2023},
  doi = {10.1103/PhysRevB.108.075156}
}

@article{Blackham2025,
  author = {Blackham, L. and Manjalingal, A. and Koshkaki, S. R. and Mandal, A.},
  title = {Microscopic theory of polaron-polariton dispersion and propagation},
  journal = {Nano Lett.},
  volume = {25},
  pages = {15874--15882},
  year = {2025},
  doi = {10.1021/acs.nanolett.5c04134}
}

@article{Raiser2017,
  author = {Raiser, D. and Mildner, S. and Ifland, B. and Sotoudeh, M. and Bl{\"o}chl, P. and Techert, S. and Jooss, C.},
  title = {Evolution of hot polaron states with a nanosecond lifetime in a manganite perovskite},
  journal = {Adv. Energy Mater.},
  volume = {7},
  pages = {1602174},
  year = {2017},
  doi = {10.1002/aenm.201602174}
}

@article{Shamblin2018,
  author = {Shamblin, J. and Heres, M. and Zhou, H. and Sangoro, J. and Lang, M. and Neuefeind, J. and Alonso, J. A. and Johnston, S.},
  title = {Experimental evidence for bipolaron condensation as a mechanism for the metal-insulator transition in rare-earth nickelates},
  journal = {Nat. Commun.},
  volume = {9},
  pages = {86},
  year = {2018},
  doi = {10.1038/s41467-017-02611-6}
}

@article{Holstein1959,
  author = {Holstein, T.},
  title = {Studies of polaron motion: Part I. The molecular-crystal model},
  journal = {Ann. Phys.},
  volume = {8},
  number = {3},
  pages = {325--342},
  year = {1959},
  doi = {10.1016/0003-4916(59)90002-8}
}

@article{Frohlich1954,
  author = {Fr{\"o}hlich, H.},
  title = {Electrons in lattice fields},
  journal = {Adv. Phys.},
  volume = {3},
  number = {11},
  pages = {325--361},
  year = {1954},
  doi = {10.1080/00018735400101213}
}

@article{Sio2019,
  author = {Sio, W. H. and Verdi, C. and Ponc{\'e}, S. and Giustino, F.},
  title = {Polarons from first principles, without supercells},
  journal = {Phys. Rev. Lett.},
  volume = {122},
  pages = {246403},
  year = {2019},
  doi = {10.1103/PhysRevLett.122.246403}
}

@article{Fratini2003,
  author = {Fratini, S. and Ciuchi, S.},
  title = {Dynamical mean-field theory of transport of small polarons},
  journal = {Phys. Rev. Lett.},
  volume = {91},
  pages = {256403},
  year = {2003},
  doi = {10.1103/PhysRevLett.91.256403}
}

@article{Wang2021,
  author = {J. Wang and D. Gao and J. Xie and W. Hu},
  title = {Polaron Hopping Induced Giant Room-Temperature Magnetodielectric Effect in Disordered Rutile {NiNb$_2$O$_6$}},
  journal = {Advanced Functional Materials},
  volume = {31},
  number = {52},
  pages = {2108950},
  year = {2021},
  doi = {10.1002/adfm.202108950}
}

@article{Wang2020,
  author = {J. Wang and D. Gao and H. Liu and J. Xie and W. Hu},
  title = {Effect of cation arrangement on polaron formation and colossal permittivity in {NiNb$_2$O$_6$}},
  journal = {Journal of Materials Chemistry C},
  volume = {8},
  number = {45},
  pages = {16107--16112},
  year = {2020},
  doi = {10.1039/D0TC04032A}
}

@misc{SupplementalMaterial,
  author = {Basu, T. and others},
  title = {Supplemental Material for ``Dispersive-phonon-driven room-temperature Ni$^{1+}$--Ni$^{2+}$ polaron hopping in spin--charge coupled rutile niobate''},
  year = {2026},
  note = {See Supplemental Material for additional experimental details, Rietveld refinement, INS ( I vs Q plot) }
}

@article{Gagne2020,
  author = {Gagn{\'e}, O. C. and Hawthorne, F. C.},
  title = {Bond-length distributions for ions bonded to oxygen: results for the transition metals and quantification of the factors underlying bond-length variation in inorganic solids},
  journal = {IUCrJ},
  volume = {7},
  number = {4},
  pages = {581--629},
  year = {2020},
  doi = {10.1107/S2052252520004673},
  url = {https://doi.org/10.1107/S2052252520004673}
}

@article{Halcrow2013,
  author = {Halcrow, M. A.},
  title = {Jahn--Teller distortions in transition metal compounds, and their importance in functional molecular and inorganic materials},
  journal = {Chemical Society Reviews},
  volume = {42},
  number = {4},
  pages = {1784--1795},
  year = {2013},
  doi = {10.1039/C2CS35253B},
  url = {https://doi.org/10.1039/C2CS35253B}
}

@article{Goodenough1955,
  author = {Goodenough, J. B.},
  title = {Theory of the Role of Covalence in the Perovskite-Type Manganites},
  journal = {Physical Review},
  volume = {100},
  pages = {564--573},
  year = {1955},
  doi = {10.1103/PhysRev.100.564},
  url = {https://doi.org/10.1103/PhysRev.100.564}
}

@misc{Basu2025,
author = {T. Basu and others},
title = {STFC ISIS Neutron and Muon Source (Inelastic Neutron Scattering Dataset)},
year = {2025},
doi = {10.5286/ISIS.E.RB2590216-1},
url = {https://doi.org/10.5286/ISIS.E.RB2590216-1}
}

@article{Kurzhals2022,
  author = {Kurzhals, P. and Kremer, G. and Jaouen, T. and Nicholson, C. W. and Heid, R. and Nagel, P. and Castellan, J. P. and Ivanov, A. and Muntwiler, M. and Rumo, M. and Salzmann, B.},
  title = {Electron-momentum dependence of electron-phonon coupling underlies dramatic phonon renormalization in YNi$_2$B$_2$C},
  journal = {Nature Communications},
  volume = {13},
  number = {1},
  pages = {228},
  year = {2022},
  doi = {10.1038/s41467-021-27821-7},
  url = {https://doi.org/10.1038/s41467-021-27821-7}
}

@article{Boca2004,
  author = {Bo{\v{c}}a, R.},
  title = {Zero-field splitting in metal complexes},
  journal = {Coordination Chemistry Reviews},
  volume = {248},
  number = {9-10},
  pages = {757--815},
  year = {2004},
  doi = {10.1016/j.ccr.2004.01.008},
  url = {https://doi.org/10.1016/j.ccr.2004.01.008}
}

@article{Sharma1966,
  author = {Sharma, R. R. and Das, T. P. and Orbach, R.},
  title = {Zero-field splitting of S-state ions. I. Point-multipole model},
  journal = {Physical Review},
  volume = {149},
  number = {1},
  pages = {257--272},
  year = {1966},
  doi = {10.1103/PhysRev.149.257},
  url = {https://doi.org/10.1103/PhysRev.149.257}
}

@article{Cheng2023,
  author = {Cheng, Y. and Wu, G. and Pajerowski, D. M. and Stone, M. B. and Savici, A. T. and Li, M. and Ramirez-Cuesta, A. J.},
  title = {Direct prediction of inelastic neutron scattering spectra from the crystal structure},
  journal = {Machine Learning: Science and Technology},
  volume = {4},
  number = {1},
  pages = {015010},
  year = {2023},
  doi = {10.1088/2632-2153/acb54d},
  url = {https://doi.org/10.1088/2632-2153/acb54d}
}

@article{Roy2026,
  title = {Quasiparticle dynamics in the $4d$-$4f$ Ising-like double perovskite Ba${}_{2}$DyRuO${}_{6}$ studied using neutron scattering and machine-learning framework},
  author = {Roy, G. and Kushwaha, E. and Kumar, M. and Ghosh, S. and Orlandi, F. and Le, M. D. and Stone, M. B. and Sannigrahi, J. and Adroja, D. T. and Basu, T.},
  journal = {Phys. Rev. B},
  year = {2026},
  month = {Apr},
  publisher = {American Physical Society},
  doi = {10.1103/dp93-x1nc},
  url = {https://link.aps.org/doi/10.1103/dp93-x1nc}
}

@article{Manti2023,
  author = {Manti, S. and Svendsen, M. K. and Kn{\o}sgaard, N. R. and Lyngby, P. M. and Thygesen, K. S.},
  title = {Exploring and machine learning structural instabilities in 2D materials},
  journal = {npj Computational Materials},
  volume = {9},
  number = {1},
  pages = {33},
  year = {2023},
  doi = {10.1038/s41524-023-00971-4},
  url = {https://doi.org/10.1038/s41524-023-00971-4}
}

@article{PhysRevB.106.195407,
  title = {Vibrational instabilities in multilayer graphene and graphite: Effects of strain and number of layers},
  author = {Zambrano Palma, Luis D. and Menezes, Marcos G. and Capaz, Rodrigo B.},
  journal = {Phys. Rev. B},
  volume = {106},
  issue = {19},
  pages = {195407},
  numpages = {7},
  year = {2022},
  month = {Nov},
  publisher = {American Physical Society},
  doi = {10.1103/PhysRevB.106.195407},
  url = {https://link.aps.org/doi/10.1103/PhysRevB.106.195407}
}

@article{Munsie2017,
  author  = {Munsie, T. J. S. and Wilson, M. N. and Millington, A. and Thompson, C. M. and Flacau, R. and Ding, C. and Guo, S. and Gong, Z. and Aczel, A. A. and Cao, H. B. and Williams, T. J.},
  title   = {Neutron diffraction and {$\mu$}SR studies of two polymorphs of nickel niobate NiNb$_2$O$_6$},
  journal = {Physical Review B},
  volume  = {96},
  number  = {14},
  pages   = {144417},
  year    = {2017},
  doi     = {10.1103/PhysRevB.96.144417}
}

@article{Maruthi2023,
  author  = {Maruthi, R. and Singh, S. and Ghosh, S. and Seehra, M. S. and Weise, B. and Prellier, W. and Thota, S.},
  title   = {Magnetic structure and field-induced transitions in the triangular spin-1 antiferromagnet NiNb$_2$O$_6$: Observation of a triple point},
  journal = {Physical Review B},
  volume  = {108},
  number  = {22},
  pages   = {224430},
  year    = {2023},
  doi     = {10.1103/PhysRevB.108.224430}
}

@article{Arnold2014,
  author       = {O. Arnold and J. C. Bilheux and J. M. Borreguero and A. Buts and
                  S. I. Campbell and L. Chapon and M. Doucet and N. Draper and
                  R. {Le Goff} and V. Lynch and A. Markvardsen and D. Mikkelson and
                  R. Mikkelson and R. Miller and K. Palmen and P. Parker and
                  G. Passos and T. Perring and P. F. Peterson and S. Ren and
                  M. A. Reuter and A. T. Savici and J. W. Taylor and
                  R. J. Taylor and R. Tolchenov and W. Zhou and J. Zikovsky},
  title        = {Mantid---Data Analysis and Visualization Package for Neutron Scattering and {$\mu$SR} Experiments},
  journal      = {Nuclear Instruments and Methods in Physics Research Section A},
  volume       = {764},
  pages        = {156--166},
  year         = {2014},
  doi          = {10.1016/j.nima.2014.07.029}
}

@article{Azuah2009,
  author       = {R. T. Azuah and L. R. Kneller and Y. Qiu and
                  P. L. W. Tregenna-Piggott and C. M. Brown and
                  J. R. D. Copley and R. M. Dimeo},
  title        = {DAVE: A Comprehensive Software Suite for the Reduction, Visualization, and Analysis of Low Energy Neutron Spectroscopic Data},
  journal      = {Journal of Research of the National Institute of Standards and Technology},
  volume       = {114},
  number       = {6},
  pages        = {341--358},
  year         = {2009},
  doi          = {10.6028/jres.114.025}
}

@article{Yang2024,
  author    = {H. Yang and C. Hu and Y. Zhou and X. Liu and Y. Shi and
               J. Li and G. Li and Z. Chen and S. Chen and C. Zeni and
               M. Horton and R. Pinsler and A. Fowler and T. Xie},
  title     = {MatterSim: A Deep Learning Atomistic Model Across Elements, Temperatures and Pressures},
  journal   = {arXiv},
  volume    = {abs/2405.04967},
  year      = {2024},
  eprint    = {2405.04967},
  archivePrefix = {arXiv},
  primaryClass = {cond-mat.mtrl-sci},
  url       = {https://arxiv.org/abs/2405.04967}
}

@article{Weber2021,
  author  = {Weber, Niklas A. and Schmidt, Hendrik and Sievert, Tim and
             Jooss, Christian and G{\"u}thoff, Friedrich and Moshneaga, Vasily and
             Samwer, Konrad and Kr{\"u}ger, Matthias and Volkert, Cynthia A.},
  title   = {Polaronic Contributions to Friction in a Manganite Thin Film},
  journal = {Advanced Science},
  volume  = {8},
  number  = {8},
  pages   = {2003524},
  year    = {2021},
  doi     = {10.1002/advs.202003524}
}

@article{Choi2015,
  author  = {Choi, W. S. and Yoo, H. K. and Ohta, H.},
  title   = {Polaron Transport and Thermoelectric Behavior in La-Doped SrTiO$_3$ Thin Films with Elemental Vacancies},
  journal = {Advanced Functional Materials},
  volume  = {25},
  number  = {5},
  pages   = {799--804},
  year    = {2015},
  doi     = {10.1002/adfm.201403023}
}

@article{Boyd2025,
  author  = {Boyd, C. and McBride, S. and Paolino, M. and Lang, M. and
             Hautier, G. and Cuk, T.},
  title   = {Assigning Surface Hole Polaron Configurations of Titanium Oxide Materials to Excited-State Optical Absorptions},
  journal = {Journal of the American Chemical Society},
  volume  = {147},
  number  = {13},
  pages   = {10981--10991},
  year    = {2025},
  doi     = {10.1021/jacs.4c15043}
}

@article{Velasco2022,
  author  = {Velasco, V. and Silva Neto, M. B. and Perali, A. and
             Wimberger, S. and Bishop, A. R. and Conradson, S. D.},
  title   = {Kuramoto synchronization of quantum tunneling polarons for describing the dynamic structure in cuprate superconductors},
  journal = {Physical Review B},
  volume  = {105},
  number  = {17},
  pages   = {174305},
  year    = {2022},
  doi     = {10.1103/PhysRevB.105.174305}
}

@article{Meggiolaro2020,
  author  = {Meggiolaro, D. and Ambrosio, F. and Mosconi, E. and
             Mahata, A. and De Angelis, F.},
  title   = {Polarons in Metal Halide Perovskites},
  journal = {Advanced Energy Materials},
  volume  = {10},
  number  = {13},
  pages   = {1902748},
  year    = {2020},
  doi     = {10.1002/aenm.201902748}
}

@article{Buizza2021,
  author  = {Buizza, L. R. and Herz, L. M.},
  title   = {Polarons and Charge Localization in Metal-Halide Semiconductors for Photovoltaic and Light-Emitting Devices},
  journal = {Advanced Materials},
  volume  = {33},
  number  = {24},
  pages   = {2007057},
  year    = {2021},
  doi     = {10.1002/adma.202007057}
}

@article{Guo2026,
  author  = {Guo, W. and Dai, J. and He, S. and Yang, L. and He, Y. and
             Duan, J. and Cen, H. and Yang, X. and Yuan, F. and
             Li, J. and Wu, Z.},
  title   = {Electron--Phonon Interactions Facilitating Large Polaron-Related Charge-Carrier Dynamics for Efficient Perovskite Nanocrystal Solar Cells},
  journal = {Advanced Science},
  volume  = {13},
  number  = {14},
  pages   = {e20934},
  year    = {2026}
}

@article{Ren2024,
  author  = {Ren, Z. and Shi, Z. and Feng, H. and Xu, Z. and Hao, W.},
  title   = {Recent Progresses of Polarons: Fundamentals and Roles in Photocatalysis and Photoelectrocatalysis},
  journal = {Advanced Science},
  volume  = {11},
  number  = {37},
  pages   = {2305139},
  year    = {2024},
  doi     = {10.1002/advs.202305139}
}

@article{Gomes2026,
  author  = {Gomes, B. M. and Holtz, J. and Pinto, M. L. and Braga, M. H.},
  title   = {Polaronic and Electrochemical Signatures in Group IVB (Ti, Zr, Hf) Oxides: Unified SKP--DFT Insights for Tunable Transport in Energy and Electronic Devices},
  journal = {Advanced Functional Materials},
  volume  = {36},
  number  = {1},
  pages   = {e09853},
  year    = {2026}
}

@article{Kong2020,
  author  = {Kong, Y. and Bo, F. and Wang, W. and Zheng, D. and Liu, H. and Zhang, G. and Rupp, R. and Xu, J.},
  title   = {Recent Progress in Lithium Niobate: Optical Damage, Defect Simulation, and On-Chip Devices},
  journal = {Advanced Materials},
  volume  = {32},
  number  = {3},
  pages   = {1806452},
  year    = {2020}
}

@article{Kampfe2016,
  author  = {K{\"a}mpfe, T. and Hau{\ss}mann, A. and Eng, L. M. and Reichenbach, P. and Thiessen, A. and Woike, T. and Steudtner, R.},
  title   = {Time-Resolved Photoluminescence Spectroscopy of Nb$^{4+}$ and O$^{-}$ Polarons in LiNbO$_3$ Single Crystals},
  journal = {Physical Review B},
  volume  = {93},
  number  = {17},
  pages   = {174116},
  year    = {2016}
}

@article{Waqar2022,
  author  = {Waqar, M. and Wu, H. and Ong, K. P. and Liu, H. and Li, C. and Yang, P. and Zang, W. and Liew, W. H. and Diao, C. and Xi, S. and Singh, D. J.},
  title   = {Origin of Giant Electric-Field-Induced Strain in Faulted Alkali Niobate Films},
  journal = {Nature Communications},
  volume  = {13},
  number  = {1},
  pages   = {3922},
  year    = {2022}
}

\subsection*{Supplemental Material}

Additional supporting information can be found online in the Supplemental Material section~\cite{SupplementalMaterial}.

\nocite{*}

\end{document}